\pdfoutput=1

\RequirePackage{amsmath}

\documentclass{llncs}

\usepackage[rgb,table,xcdraw]{xcolor}

\definecolor{diff-red}{RGB}{255,64,64}
\definecolor{diff-green}{RGB}{0,160,0}
\definecolor{diff-blue}{RGB}{0,0,255}
\definecolor{diff-orange}{RGB}{255,128,48}

\definecolor{diff-0-of-25}{RGB}{99,190,123}
\definecolor{diff-1-of-25}{RGB}{176,212,127}
\definecolor{diff-2-of-25}{RGB}{255,249,135}
\definecolor{diff-3-of-25}{RGB}{255,230,132}
\definecolor{diff-4-of-25}{RGB}{255,220,125}
\definecolor{diff-5-of-25}{RGB}{251,210,117}

\definecolor{diff-10-of-25}{RGB}{251,222,139}
\definecolor{diff-11-of-25}{RGB}{252,214,135}
\definecolor{diff-12-of-25}{RGB}{253,206,131}
\definecolor{diff-13-of-25}{RGB}{254,198,127}
\definecolor{diff-14-of-25}{RGB}{255,190,123}

\definecolor{diff-16-of-25}{RGB}{255,166,120}
\definecolor{diff-17-of-25}{RGB}{250,156,115}
\definecolor{diff-18-of-25}{RGB}{245,146,110}
\definecolor{diff-19-of-25}{RGB}{240,136,105}

\definecolor{diff-24-of-25}{RGB}{240,110,103}
\definecolor{diff-25-of-25}{RGB}{252,104,103}

\usepackage{bm}
\newcommand{\difftilde}[0]{\raisebox{-.4\height}{\hbox{\char`~}}}
\newcommand{\diffplus}[0]{\raisebox{.3\height}{\scalebox{.7}{+}}}
\newcommand{\diffminus}[0]{\raisebox{.4\height}{\text{-}}}
\newcommand{\diffabsent}[0]{\boldsymbol{-}}

\usepackage{tikz}
\usetikzlibrary{automata, calc, positioning, arrows, arrows.meta, shapes, shapes.geometric}
\usepackage{pgfplots}
\pgfplotsset{compat=1.17}

\usepackage[caption=false]{subfig}

\usepackage{wrapfig}

\usepackage[shortlabels]{enumitem}

\usepackage{amsfonts}

\usepackage{array}
\usepackage{booktabs}
\usepackage{rotating}
\newcolumntype{P}[2]{%
    >{\begin{turn}{#1}\begin{minipage}{#2}\small\raggedright\hspace{0pt}}l%
    <{\end{minipage}\end{turn}}%
}

\newcommand{\tikzmark}[1]{\tikz[remember picture,overlay] \node [] (#1) {};}

\PassOptionsToPackage{hyphens}{url}\usepackage[hidelinks]{hyperref}
\makeatletter
\AtBeginDocument{%
    \def\doi#1{\url{https://doi.org/#1}}}
\makeatother

\makeatletter
\newcommand\notsotiny{\@setfontsize\notsotiny\@vipt\@viipt}
\makeatother

\title{A Multi-level Methodology for Behavioral Comparison of Software-Intensive Systems}

\author{
    Dennis Hendriks\inst{1,2}\thanks{This research is carried out as part of the Transposition project under the responsibility of ESI (TNO) in co-operation with ASML. The research activities are supported by the Netherlands Ministry of Economic Affairs and TKI-HTSM.} \and
    Arjan van der Meer\inst{1,3}$^*$ \and
    Wytse Oortwijn\inst{1}$^*$
}
\institute{
    ESI (TNO), Eindhoven, The Netherlands\\
    \email{dennis.hendriks@tno.nl} \and
    Radboud University, Nijmegen, The Netherlands\\
    \email{dennis.hendriks@ru.nl} \and
    Capgemini Engineering, Eindhoven, The Netherlands
}

\begin{document}

\maketitle
\begin{abstract}
Software-intensive systems constantly evolve.
To prevent software changes from unintentionally introducing costly system defects, it is important to understand their impact to reduce risk.
However, it is in practice nearly impossible to foresee the full impact of software changes when dealing with huge industrial systems with many configurations and usage scenarios.
To assist developers with change impact analysis we introduce a novel multi-level methodology for behavioral comparison of software-intensive systems.
Our fully automated methodology is based on comparing state machine models of software behavior.
We combine existing complementary comparison methods into a novel approach, guiding users step by step though relevant differences by gradually zooming into more and more detail.
We empirically evaluate our work through a qualitative exploratory field study, showing its practical value using multiple case studies at ASML, a leading company in developing lithography systems.
Our method shows great potential for preventing regressions in system behavior for software changes.


\end{abstract}

\section{Introduction}
\label{sec:introduction}

Software-intensive systems, e.g., cyber-physical systems, become more and more complex.
They often employ a component-based software architecture to manage their complexity.
Over the years such systems continuously evolve by adding new features and addressing defects, more and more layers are built on top of each other~\cite{Klusener2018}, and components that are not well-maintained become legacy~\cite{Lehman1980,Schuts2016}.

Changing the software is often considered risky as any change can potentially break a system.
If a software change leads to a system defect, then the impact can be tremendous due to system downtime and productivity loss~\cite{Schuts2016}.
This may even lead to software engineers becoming afraid to make changes for which they can't properly foresee the impact on (other parts of) the system.

To reduce the risks, it is essential to understand the impact of software changes.
However, for large complex industrial code bases consisting of tens of millions of lines of code, no single person has the complete overview.
This makes it difficult to understand the impact of software changes on the overall system functionality~\cite{Gulzar2019}.
This is especially true when the system can behave differently for different configurations and usage scenarios~\cite{Yang2021}.

It is thus important that:
1) software developers understand how the system currently behaves for different configurations and usage scenarios, and
2) they understand how software changes impact that system behavior.

To address these needs, in this paper we introduce a novel multi-level methodology for behavioral comparison of (large) software-intensive systems.
The power of our methodology is that it quickly guides users to relevant differences.
This avoids the laborious and error-prone practice of looking into many thousands of lines of code, or plough through gigabytes of execution logs.
Our method is fully automated, making it possible to consider huge (sub-)systems, for which due to their sheer size it is practically impossible to compare their behavior manually.

Our methodology is based on comparing state machine models rather than source code or execution logs, which makes it generally applicable.
State machines can compactly and intuitively represent system behavior as a collection of software function calls and the order in which they are called.
Such models are general and can be obtained by any means of model learning or construction.

Methods to compare state machines can be divided into two classes that complement each other~\cite{Walkinshaw2013}.
Language-based methods compare state machines in terms of their allowed sequences of function calls, while structure-based methods compare them in terms of their states and transitions.

However, two important things are missing in literature:
1) a single automated method integrating these individual methods to allow large-scale industrial application, and
2) an approach to inspect the resulting differences at various levels of detail, and step by step zoom in on relevant differences, to manage the complexity of huge systems.
Our methodology tackles both these challenges.

Our methodology takes any number of sets of state machines representing software behavior of, e.g., different software versions, different configurations or different usage scenarios.
We automatically compare the provided sets by comparing the languages and structures of their state-machine models.
The comparison results can be inspected at six levels of abstraction, ranging from very high-level differences to very detailed ones.
Users are guided through the differences in a step by step fashion tailored to allow them to zoom in on relevant behavioral differences, wasting no time on irrelevant ones.

We empirically evaluate the practical potential of our methodology through a qualitative exploratory field study~\cite{Runeson2009,Storey2020}.
Using multiple case studies at ASML, a leading company in developing lithography systems,
we demonstrate that our approach can be applied to large industrial (sub-)systems, provides developers and architects insight into their behavioral differences, and allows them to
find unintended regressions.
The company wants to broadly adopt our work.

The remainder of this paper is organized as follows.
In Section~\ref{sec:background} we introduce the concepts, definitions and methods on which we build our methodology.
Section~\ref{sec:methodology} introduces our methodology, both conceptually and formally.
We evaluate our methodology in Section~\ref{sec:evaluation}, before concluding in Section~\ref{sec:conclusins-and-future-work}.

\vspace{-8pt}
\section{Background}
\label{sec:background}
\vspace{-2pt}

\vspace{-4pt}
\subsection{Software Behavior}
\vspace{-2pt}

Programming languages typically have a notion of \emph{function}, \emph{procedure} or \emph{method}.
The behavior of software implemented in such languages can then be seen as all the \emph{calls} to or \emph{invocations} of these functions, and the constraints on the order in which they may be called.

Large systems often employ a component-based software architecture to manage their complexity.
The many components are independent units of development and deployment, encapsulate functionality and allow for re-use~\cite{McIlroy1968,Szyperski2002,Vitharana2003}.
Functions may then be called internally within a component and to communicate between components connected via interfaces, e.g., remote procedure calls.

\vspace{-10pt}
\subsection{State Machines}
\label{sub:sw-behavior-as-state-machines}
\vspace{-2pt}

We consider software behavior in terms of sequences of discrete \emph{events}, e.g., the start and end of function calls.
We define an \emph{alphabet} $\Sigma$ to be a finite set of events of interest.
A \emph{trace} $t\in\Sigma^*$ represents a single finite execution, with $*$ the Kleene star.
The length of $t$ is denoted by $|t|$ and its $i$-th event by $t_i$ for $1\leq i\leq |t|$.
An execution \emph{log} is a set of observed traces, and can for instance be obtained by explicit logging or through sniffing tools.

A \emph{state machine} or \emph{automaton} compactly and intuitively represents multiple executions.
We define a \emph{Non-deterministic Finite Automaton} (NFA) $A=$ \linebreak$(S,\Sigma,\Delta,I,F)$ as a 5-tuple, with $S$ a finite set of states, $\Sigma$ a finite set of events (the alphabet), $\Delta\subseteq S\times\Sigma\times S$ a set of transitions, $I\subseteq S$ a set of \emph{initial states}, and $F\subseteq S$ a set of \emph{accepting states}.
\emph{Deterministic Finite Automata} (DFAs) are a sub-class of NFAs allowing for each source state and event only a single target state.
An NFA can be determinized to a DFA~\cite{Glabbeek2008}.

A trace $t\in\Sigma^*$ is \emph{accepted} by an NFA $A=(S,\Sigma,\Delta,I,F)$ iff there exists a sequence $(s_0,t_1,s_1),(s_1,t_2,s_2),...\,,(s_{|t|-1}, t_{|t|}, s_{|t|})\in\Delta^*$ with $s_0\in I$ and $s_{|t|}\in F$.
Traces that are not accepted are \emph{rejected}.
The \emph{language} $\mathcal{L}(A)$ of an NFA $A$ is the set of all its accepted traces, i.e., $\mathcal{L}(A)=\{t\in\Sigma^*\,|\,A~\textrm{accepts}~t\}$.
The behavior presence predicate $B(A)$ indicates whether $A$ has any behavior, i.e., $B(A)=\allowbreak(\mathcal{L}(A)\neq\emptyset)$.
State machines can be \emph{minimized} to a representation with the least number of states possible, while still accepting the same language~\cite{Hopcroft1971,Paige1987}.
Given two NFAs $A_1$ and $A_2$, union and intersection are defined as operations that reflect the effect on their resulting languages, i.e., $\mathcal{L}(A_1\cup A_2)=\mathcal{L}(A_1)\cup\mathcal{L}(A_2)$ and $\mathcal{L}(A_1\cap A_2)=\mathcal{L}(A_1)\cap\mathcal{L}(A_2)$, respectively~\cite{Sipser2013}.

A (minimal) state machine can be obtained from an execution log through model learning, e.g., using state machine learning algorithms~\cite{Gold1967,Higuera2010,Lang1998,Hooimeijer2022} or through active automata learning~\cite{Higuera2010,Howar2018}.
Their details are beyond the scope of this paper.

\begin{figure}[!b]
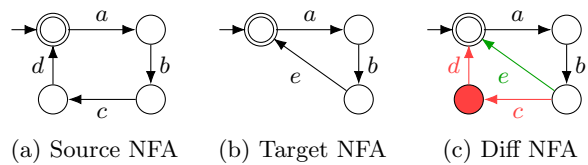

    \vspace{-11pt}
    \centering
    \subfloat[Source NFA]{
        \begin{tikzpicture}[scale=1.0,]
            \input{figures/2-diff-state-machine-src.tex}
        \end{tikzpicture}
        \label{fig:2-diff-state-machine-src}
    }
    \quad
    \subfloat[Target NFA]{
        \begin{tikzpicture}[scale=1.0]
            \input{figures/2-diff-state-machine-tgt.tex}
        \end{tikzpicture}
        \label{fig:2-diff-state-machine-tgt}
    }
    \quad
    \subfloat[Diff NFA]{
        \begin{tikzpicture}[scale=1.0]
            \input{figures/2-diff-state-machine-diff.tex}
        \end{tikzpicture}
        \label{fig:2-diff-state-machine-diff}
    }
    \vspace{-6pt}
    \caption{Source and target NFAs and their structural differences as a diff NFA.}
    \label{fig:2-diff-state-machine}
    \vspace{-12pt}
\end{figure}

\vspace{-20pt}
\subsection{State Machine Comparison}
\label{sub:state-machine-comparison}
\vspace{-4pt}

There are various ways to compare state machines.
Walkinshaw et al. differentiate two perspectives: language-based and structure-based comparisons~\cite{Walkinshaw2013}.

The language perspective considers to which extend the languages of state machines overlap.
Two state machines $A_1$, $A_2$ are \emph{language equivalent} ($=_L$) iff they accept exactly the same language, i.e., $A_1=_L A_2 \Leftrightarrow \mathcal{L}(A_1)=\mathcal{L}(A_2)$.
A state machine $A_1$ is related by \emph{language inclusion} ($\leq_L$) to state machine $A_2$ iff the language of $A_1$ is included in that of $A_2$, i.e., $A_1\leq_L A_2 \Leftrightarrow \mathcal{L}(A_1)\subseteq\mathcal{L}(A_2)$.
Various other types of well-known binary equivalence and inclusion relations exist~\cite{Glabbeek1993}, as well as non-binary ones such as precision and recall~\cite{Sokolova2009,Walkinshaw2013}.
We use language equivalence and inclusion as these are commonly used in automata theory, are sufficient to capture the order of function calls, and can be easily explained even to engineers without a formal background.
For finite state machines these relations can be computed on their finite structures~\cite{Cleaveland2001}.

Language-based comparison considers the externally observable behavior of state machines.
Complementary to it, structure-based comparison considers the overlap of their internal representations in terms of states and transitions.

Walkinshaw et al. define the \emph{LTSDiff} algorithm~\cite{Walkinshaw2013} that
takes two state machines and computes a \emph{diff} state machine: a compact representation of their differences.
An example is shown in Figure~\ref{fig:2-diff-state-machine}.
A diff state machine is a regular state machine with its states and transitions annotated to represent difference information, i.e. `unchanged' (black), `added' (green) and `removed' (red).

The algorithm has three steps:
1) Compute similarity scores for all possible pair-wise combinations of states from the two NFAs being compared.
A local score considers only the overlap in directly connected incoming and outgoing transitions of the states.
It is extended to a global score by recursively considering all context, using an attenuation factor to ensure closer-by context counts more towards the score than further away context.
2) Use the scores to heuristically compute a matching between states of the two NFAs based on landmarks, a percentage of the highest scoring pairs that score at least some factor better than any other pairs, with a fallback to the initial states.
The most obviously equivalent state pairs are matched first and these are then used to match the surrounding areas, rejecting any remaining conflicting state pairs.
The next-best remaining state pair is then selected and matched, etc, until no state pairs are left to consider.
3) Use the matching to compute the diff state machine.

The LTSDiff algorithm has the advantage that it does not require states to be reachable from initial states, does not require state machines to be deterministic or minimal, does not rely on state labels, and that it produces relatively small diffs in practice, unlike some other approaches~\cite{Sokolsky2006,Nejati2007,Quante2007,Kelter2008}.

For a more extensive overview of alternative approaches to compare the language and structure of state machines, see the work of Walkinshaw et al.~\cite{Walkinshaw2013}.

\vspace{-18pt}
\section{Behavioral Comparison Methodology}
\label{sec:methodology}
\vspace{-5pt}

The language and structure-based state machine comparison approaches are complementary.
However, to the best of our knowledge there is no work that fully exploits the complementary nature of these approaches, to provide intuitive insights into the behavioral impact of changes for industrial-scale software-intensive systems.
Our methodology takes advantage of their complementary nature in a novel way, to allow handling the complexity of such scale.

\begin{figure}[!b]
    \centering
    \begin{tikzpicture}[scale=1.0,]
        \input{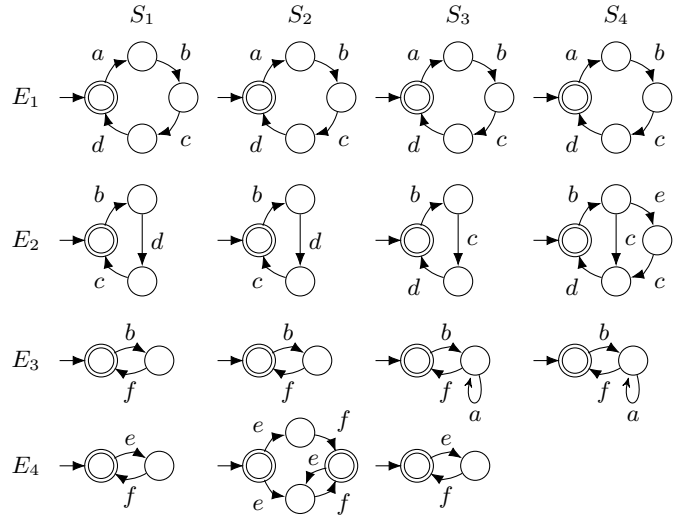}
    \end{tikzpicture}
    \vspace{-10pt}
    \caption{The input state machines for the running example, for entities $E_1$ through $E_4$ (rows) and model sets $S_1$ through $S_4$ (columns). $S_4(E_4)=(\emptyset,\emptyset,\emptyset,\emptyset,\emptyset)$.}
    \label{fig:3-running-example-input}
    \vspace{-18pt}
\end{figure}

As input our methodology takes any number of \emph{model sets} representing, e.g., different software versions, configurations or usage scenarios.
They contain state machines that represent behaviors of a number of \emph{entities} representing, e.g., software functions or components.
Formally, let $E$ be a finite set of (behavioral) entities and $\mathcal{N}$ the set of all NFAs.
A model set $S\in E\rightarrow \mathcal{N}$ is a complete mapping of entities to models (NFAs).
An incomplete mapping can be made complete using $(\emptyset,\emptyset,\emptyset,\emptyset,\emptyset)$ as NFA for unmapped entities.
As input our methodology takes a finite entities set $E$ and a finite set of model sets $\mathbb{S}=\{S_1,...,S_n\}\subseteq E\rightarrow \mathcal{N}$.

Figure~\ref{fig:3-running-example-input} shows the model sets that we use as a running example.
For model set $S_4$ (e.g., configuration 4) there is no model for entity $E_4$ (e.g., function 4).
If these models were obtained through model learning on execution logs, no behavior was observed for function 4 using configuration 4.

\begin{figure}[!t]
    \vspace{-8pt}
    \centering
    \begingroup
    \arrayrulecolor{white}
    \setlength\arrayrulewidth{2pt}
    \setlength{\tabcolsep}{7pt}
    \def\arraystretch{1.2}
    \small
    \begin{tabular}{c | c | c | c | c | c}
        \arrayrulecolor{white}
        \rowcolor{black!10}
        \multicolumn{3}{c|}{Model sets} & \multicolumn{3}{c}{Models} \\

        \hline
        \rowcolor{black!10}
        Level 1 & Level 2 & Level 3 & Level 4 & Level 5 & Level 6 \\

        \rowcolor{black!10}
        Variants & Variant & Variant & Variants & Variant & Variant \\

        \rowcolor{black!10}
        ~ & relations & differences & ~ & relations & differences \\

        \hline
        \rowcolor{black!10}
        L & L & L & L & L / S & S \\
    \end{tabular}
    \endgroup
    \vspace{-3pt}
    \caption{Methodology overview: six levels of detail to inspect comparison results.}
    \label{fig:3-behavioral-comparison-levels-overview}
    \vspace{-14pt}
\end{figure}

Our methodology compares the states machines of all input model sets.
The results are represented at six levels of abstraction (Figure~\ref{fig:3-behavioral-comparison-levels-overview}).
The first three levels focus on model sets and the last three on individual (models of) entities within them.
For both model sets and models, the first level considers different behavioral variants, the second level relates the variants, and the third level elaborates on variant differences.
Users are guided step by step through the levels, by gradually zooming in to more detail, letting them focus on relevant differences.
Levels 1\,--\,5 contain information from the language perspective (L), while levels 5 and 6 contain information from the structural perspective (S).
Next, we further elaborate on each of the six levels.

\vspace{-8pt}
\subsection{Level 1: Model Set Variants}
\label{sub:level1}
\vspace{-3pt}

Level 1 provides the highest level overview.
It shows whether model sets have the same behavior, i.e., their entity models are language equivalent.
Two model sets $S_i, S_j\in\mathbb{S}$ have the same behavior, denoted $S_i =_L S_j$, iff $\forall_{e\in E}\,S_i(e) =_L S_j(e)$.

We compare model sets against each other and determine unique model set behavior variants.
Variants are formally defined to be equivalence classes of $\mathbb{S}$ under $=_L$, so that $\mathbb{S}/\hspace{-0.4em}=_L$ is the set of all variants. For presentational clarity we enumerate and refer to different variants of $\mathbb{S}$ in alphabetical order: A, B, etc.
We choose a structural representative for each behavioral equivalence class.

\begin{figure}[!t]
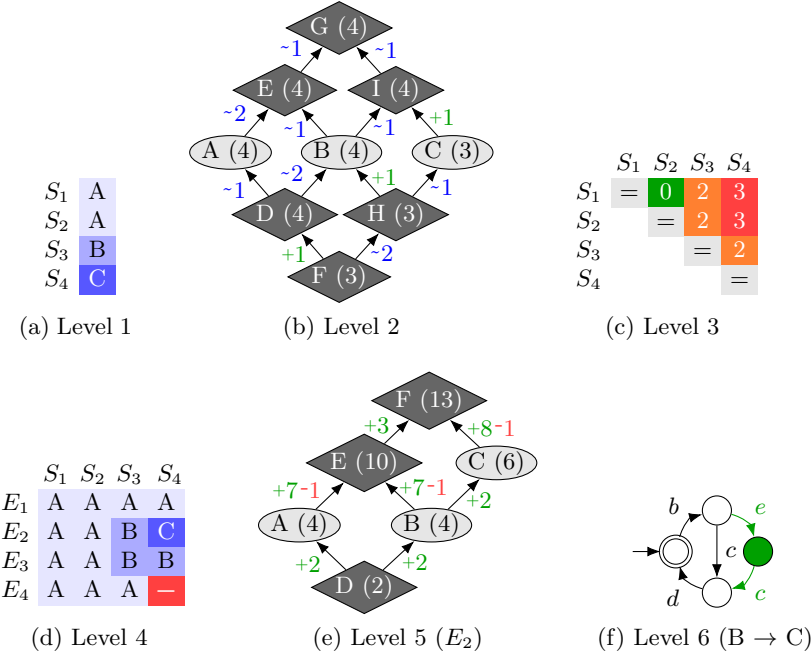

    \vspace{-10pt}
    \centering
    \subfloat[Level 1]{
        \begin{tikzpicture}[scale=1.5]
            \tikzset{
  ->,>=stealth',auto,node distance=22pt,
  arrows={-{Latex[scale=1.05]}},
  every initial by arrow/.style={-{Latex[scale=1.05]}},
  every initial by arrow/.append style={anchor/.append style={shape=coordinate}},
  every node/.style={font=\small},
  every state/.style={circle, draw, minimum size=11pt, initial text=},
  initial distance=10pt,
  initial where=left,
  accepting/.style={double, double distance=1pt},
  background rectangle/.style={draw=red, top color=blue, rounded corners},
  obsvariant/.style ={state,ellipse,         fill=black!10,inner sep=0pt,minimum size=0pt,align=center,font=\footnotesize},
  calcvariant/.style={state,diamond,aspect=2,fill=black!60,inner sep=0pt,minimum size=0pt,align=center,font=\footnotesize,text=white},
  myedge/.style={}
}

\node[] {
    \begingroup
    \begin{tabular}{ >{\centering\arraybackslash}m{1.5em} *{1}{>{\centering\arraybackslash}m{1.2em}} }
        $S_1$ & \cellcolor{blue!10}A              \\
        $S_2$ & \cellcolor{blue!10}A              \\
        $S_3$ & \cellcolor{blue!33}B              \\
        $S_4$ & \cellcolor{blue!66}\color{white}C \\
    \end{tabular}
    \endgroup
};

        \end{tikzpicture}
        \label{fig:3-running-example-output-lvl1}
    }
    \qquad
    \subfloat[Level 2]{
        \begin{tikzpicture}[scale=1.5]
            \input{figures/3-running-example-output-lvl2.tex}
        \end{tikzpicture}
        \label{fig:3-running-example-output-lvl2}
    }
    \qquad
    \subfloat[Level 3]{
        \begin{tikzpicture}[scale=1.5]
            \tikzset{
  ->,>=stealth',auto,node distance=22pt,
  arrows={-{Latex[scale=1.05]}},
  every initial by arrow/.style={-{Latex[scale=1.05]}},
  every initial by arrow/.append style={anchor/.append style={shape=coordinate}},
  every node/.style={font=\small},
  every state/.style={circle, draw, minimum size=11pt, initial text=},
  initial distance=10pt,
  initial where=left,
  accepting/.style={double, double distance=1pt},
  background rectangle/.style={draw=red, top color=blue, rounded corners},
  obsvariant/.style ={state,ellipse,         fill=black!10,inner sep=0pt,minimum size=0pt,align=center,font=\footnotesize},
  calcvariant/.style={state,diamond,aspect=2,fill=black!60,inner sep=0pt,minimum size=0pt,align=center,font=\footnotesize,text=white},
  myedge/.style={}
}

\node[] {
    \begingroup
    \begin{tabular}{ >{\centering\arraybackslash}m{1.5em} *{4}{>{\centering\arraybackslash}m{1.2em}}  }
        ~     & $S_1$                   & $S_2$                                  & $S_3$                                   & $S_4$                                   \\
        $S_1$ & \cellcolor{black!10}$=$ & \cellcolor{diff-green}\color{white}$0$ & \cellcolor{diff-orange}\color{white}$2$ & \cellcolor{diff-red}\color{white}$3$    \\
        $S_2$ &                         & \cellcolor{black!10}$=$                & \cellcolor{diff-orange}\color{white}$2$ & \cellcolor{diff-red}\color{white}$3$    \\
        $S_3$ &                         &                                        & \cellcolor{black!10}$=$                 & \cellcolor{diff-orange}\color{white}$2$ \\
        $S_4$ &                         &                                        &                                         & \cellcolor{black!10}$=$                 \\
    \end{tabular}
    \endgroup
};

        \end{tikzpicture}
        \label{fig:3-running-example-output-lvl3}
    }
    \\
    \subfloat[Level 4]{
        \begin{tikzpicture}[scale=1.5]
            \tikzset{
  ->,>=stealth',auto,node distance=22pt,
  arrows={-{Latex[scale=1.05]}},
  every initial by arrow/.style={-{Latex[scale=1.05]}},
  every initial by arrow/.append style={anchor/.append style={shape=coordinate}},
  every node/.style={font=\small},
  every state/.style={circle, draw, minimum size=11pt, initial text=},
  initial distance=10pt,
  initial where=left,
  accepting/.style={double, double distance=1pt},
  background rectangle/.style={draw=red, top color=blue, rounded corners},
  obsvariant/.style ={state,ellipse,         fill=black!10,inner sep=0pt,minimum size=0pt,align=center,font=\footnotesize},
  calcvariant/.style={state,diamond,aspect=2,fill=black!60,inner sep=0pt,minimum size=0pt,align=center,font=\footnotesize,text=white},
  myedge/.style={}
}

\node[] {
    \begingroup
    \begin{tabular}{ >{\centering\arraybackslash}m{1.5em} *{4}{>{\centering\arraybackslash}m{1.2em}}  }
        ~     & $S_1$                & $S_2$                & $S_3$                & ~$S_4$~                                        \\
        $E_1$ & \cellcolor{blue!10}A & \cellcolor{blue!10}A & \cellcolor{blue!10}A & \cellcolor{blue!10}A                           \\
        $E_2$ & \cellcolor{blue!10}A & \cellcolor{blue!10}A & \cellcolor{blue!33}B & \cellcolor{blue!66}\color{white}C              \\
        $E_3$ & \cellcolor{blue!10}A & \cellcolor{blue!10}A & \cellcolor{blue!33}B & \cellcolor{blue!33}B                           \\
        $E_4$ & \cellcolor{blue!10}A & \cellcolor{blue!10}A & \cellcolor{blue!10}A & \cellcolor{diff-red}\color{white}$\diffabsent$ \\
    \end{tabular}
    \endgroup
};

        \end{tikzpicture}
        \label{fig:3-running-example-output-lvl4}
    }
    \qquad
    \subfloat[Level 5 ($E_2$)]{
        \begin{tikzpicture}[scale=1.5]
            \input{figures/3-running-example-output-lvl5.tex}
        \end{tikzpicture}
        \label{fig:3-running-example-output-lvl5}
    }
    \qquad
    \subfloat[Level 6 (B $\rightarrow$ C)]{
        \begin{tikzpicture}[scale=1.5]
            \input{figures/3-running-example-output-lvl6.tex}
        \end{tikzpicture}
        \label{fig:3-running-example-output-lvl6}
    }
    \vspace{-4pt}
    \caption{Behavioral comparison methodology output for the running example: complete levels 1\,--\,4, level 5 for $E_2$, and level 6 for $E_2$ variants B $\rightarrow$ C.}
    \label{fig:3-running-example-output}
    \vspace{-14pt}
\end{figure}

Figure~\ref{fig:3-running-example-output-lvl1} shows the level 1 result for our running example.
Model sets $S_1$ and $S_2$ have the same behavior for all four functions and thus get variant A, even though their models for $E_4$ are structurally different.
Model sets $S_3$ and $S_4$ get variants B and C as they differ from the other model sets (and each other).

Level 1 thus provides a very high level overview of which model sets have the same or different behavior, and how few or many variants there are.
We can see whether this matches our expectations.
Depending on the use case, we may be satisfied already after looking at these results.
For instance, if we want to know whether different configurations have the same behavior, and if they all have the same variant, we can already conclude that there are no differences in their behavior.
If we do go to the other levels, we can ignore model set $S_2$ as it has the same behavior as $S_1$.
In fact, from the language perspective we can focus on (representatives of) model set variants, each representing one or more models with the same behavior, rather than on individual model sets.
Finally, in Figure~\ref{fig:3-running-example-output-lvl1} variants are colored using shades of blue like a heat map.
In case of many model sets this may reveal patterns, as we will see in Section~\ref{sec:evaluation}.

\vspace{-8pt}
\subsection{Level 2: Model Set Variant Relations}
\label{sub:level2}
\vspace{-3pt}

Level 1 provides us with model set variants that each have different behavior.
Level 2 provides more details.
It considers whether the behavior of one model set variant is completely included in the behavior of another variant, i.e., it has less behavior.
Formally, for two model sets $S_i, S_j\in\mathbb{S}$, $S_i$ is related to $S_j$ by language inclusion, denoted $S_i\leq_L S_j$, iff $\forall_{e\in E}\,S_i(e)\leq_L S_j(e)$.
Given that all model set variants have different behavior, $S_i$ thus has less behavior for at least one entity.
Partially ordered set $(\mathbb{S}/\hspace{-0.4em}=_L,\leq_L)$ can be extended into a finite lattice by computing unions (as supremum) and intersections (as infimum) of representatives of model set variants until a fixed point is reached.
The union or intersection of two model sets constitutes the per-entity pairwise combination of their entity models, using state machine union or intersection, respectively.

Figure~\ref{fig:3-running-example-output-lvl2} shows the level 2 lattice for our running example.
The variants from level 1 are indicated by ellipses containing the variant and number of entity models that have behavior.
The extra variants computed to complete the lattice are indicated by diamonds.
Arrows indicate inclusion relations, e.g., the behavior of variant D is included in that of variants A and B (and E, I and G, by transitivity).
The arrows are labeled with the number of entities with different present behavior (e.g., {\color{diff-blue}$\difftilde 1$}) and the number of entities with newly present behavior (e.g., {\color{diff-green}$\diffplus 1$}).
Formally, for model set variants $S_i,S_j$ and $S_i\leq_L S_j$, these are computed by $|\{e\in E\,|\,B(S_i(e))\land B(S_j(e))\land S_i(e)\neq_L S_j(e)\}|$ and $|\{e\in E\,|\,\lnot B(S_i(e))\land B(S_j(e))\}|$, respectively.

Level 2 provides information on which variants have more or less behavior than other variants, whether variants are closely related (direct arrow) or less closely related (via several arrows), and it has quantitative information on the models within the model sets by means of the labels on the arrows.
As for level 1, we can check whether this conforms to our expectations, or not.
For instance, if we compare two software versions and we only added new functionality (e.g., new entities), we would reasonably expect the behavior of the old software version to be included in that of the new software version, and we can check whether that is indeed the case.
If this is all that we want to know, we can stop here and we don't need to proceed to level 3.

\vspace{-9pt}
\subsection{Level 3: Model Set Variant Differences}
\label{sub:level3}
\vspace{-3pt}

Level 2 shows us the quantitative differences between model sets via the arrow labels.
However, some model set variants are not directly related by an inclusion arrow (e.g., variants A and B).
The number of entities with different behavior between them can't be determined from the lattice, as simply summing labels (e.g., {\color{diff-blue}$\difftilde 1$}, {\color{diff-green}$\diffplus 1$}) could count the same entity multiple times.
Level 3 provides more details, showing the number of entities with different behavior between all input model sets. That is, for model sets $S_i,S_j\in\mathbb{S}$ it shows $|\{e\in E\,|\,S_i(e)\neq_L S_j(e)\}|$.

Figure~\ref{fig:3-running-example-output-lvl3} shows the level 3 matrix for our running example.
Rows and columns are labeled with the input model sets.
Cells indicate the number of entities with different behavior.
As language (in)equality is a symmetric and reflexive relation, only the upper-right part of the matrix is filled, and the diagonal is labeled with `$=$' symbols.
As expected, model sets $S_1$ and $S_2$ have zero entities with different behavior, as they have the same model set variant.
Model sets $S_1$ (variant A) and $S_4$ (variant C) have three entities with different behavior.

Level 3 provides more detailed quantitative information.
It shows not just whether model sets are different, and how many model sets have differences, but also how different they are.
The diagonal is colored gray as it is not relevant.
Numbered cells are colored like a heat map based on a gradient from green (no entities with differences) via yellow and orange to red (most entities with differences).
In case of many model sets this may again reveal patterns{, as we will see in Section~\ref{sec:evaluation}.
Similarly to the previous levels, we can check whether all information matches our expectations, and whether we want to proceed to level 4, or not.

\vspace{-9pt}
\subsection{Level 4: Model Variants}
\label{sub:level4}
\vspace{-3pt}

Levels 1\,--\,3 focus on model sets.
Level 4 zooms in even further and considers the (entity) models within the model sets.
Similar to how level 1 identifies model set variants, level 4 identifies model variants for each entity.
Formally, for an entity $e\in E$, let $\mathbb{S}_e=\{S(e)\,|\,S\in\mathbb{S}\}$.
We consider equivalence classes $\mathbb{S}_e/\hspace{-0.4em}=_L$ for each $e\in E$ and enumerate and represent them in alphabetical order: A, B, etc.
Note that variants are determined per entity and thus variant A of one entity does not necessarily have the same behavior as variant A of another entity.

Figure~\ref{fig:3-running-example-output-lvl4} shows the level 4 matrix for our running example.
The cells indicate the behavior variant of the model for the corresponding entity (row) in the corresponding model set (column).

Level 4 is the first level to provide details on which entities differ between model sets.
This provides a high level overview of the behavior variants for entity models, similar to how level 1 provides it for model sets.
We can see the variants, how many there are, for which models sets, and whether this is expected or not.
Depending on the use case, we may again stop at this level if it answers our questions, e.g., in case of checking for regressions if each entity has only a single behavior variant.
Otherwise, we can reduce the number of entities to consider for subsequent levels, e.g., skip the ones without regressions (only a single variant, no differences).
Furthermore, we may then focus only on unique entity model variants instead of all individual entity models.
Finally, the matrix cells are again colored using shades of blue like a heat map.
Models without behavior are indicated as a red cell labeled `$\diffabsent$' to make them stand out.
Here too, in case of many model sets this may reveal patterns, as we will see in Section~\ref{sec:evaluation}.

\vspace{-10pt}
\subsection{Level 5: Model Variant Relations}
\label{sub:level5}
\vspace{-3pt}

Level 5 shows relations between entity model variants of level 4, similar to how level 2 shows relations between model set variants of level 1.
Formally, for an entity $e\in E$ we have a partially ordered set $(\mathbb{S}_e/\hspace{-0.4em}=_L,\leq_L)$, which we extend to a finite lattice using unions and intersections, similar to level 2.

Figure~\ref{fig:3-running-example-output-lvl5} shows the level 5 lattice for our running example, for entity $E_2$.
We use a representative model for each entity model variant (set of equivalent models).
The node shapes and arrows are as in level 2.
The node labels now indicate the number of transitions of the model, and the arrow labels indicate the number of added (e.g., {\color{diff-green}$\diffplus 7$}) and removed transitions (e.g., {\color{diff-red}$\diffminus 1$}).
These are based on the structural comparison that we use and will explain further for level 6.
In our example, the behavior of variant B is included in the behavior of variant C.

Level 5 provides information on which entity model variants have more or less behavior, how closely they are related, and the amount of changes between them.
As for previous levels, we can check whether this conforms to our expectations, or not.
We can also use it to decide what to inspect in more detail in level 6.

\vspace{-10pt}
\subsection{Level 6: Model Variant Differences}
\label{sub:level6}
\vspace{-3pt}

Level 6 is the last level.
It shows all structural differences between two entity model variants of level 5 as a diff NFA, computed with the LTSDiff algorithm.

Figure~\ref{fig:3-running-example-output-lvl6} shows the level 6 diff NFA for our running example, for variants B and C of entity $E_2$.
Variant C (from model set $S_4$) has two extra transitions in its state machine, and this is clearly visible as two green arrows in this figure.

Level 6 provides the most detailed behavioral differences.
Diff NFAs show differences in terms of states and transitions within models.
As with the other levels, we can check whether this matches our expectations, or not.

\vspace{-5pt}
\section{Evaluation}
\label{sec:evaluation}
\vspace{-3pt}

We perform an empirical evaluation of our methodology through an exploratory field study~\cite{Runeson2009,Storey2020}.
To gain some first evidence of both its practical potential and its ability to handle large systems, we perform three case studies at ASML.
The first two case studies provide some preliminary evidence of our methodology's practical value, by showing the benefits of all six of its levels, as well as finding a regression.
The third case study shows that our methodology can be applied to a large industrial system, providing insights into its behavior.
We have completely automated our approach, in a (for now) company-internal prototype tool.

ASML develops photolithography systems for the semiconductor industry.
These systems process \emph{wafers} (thin circular slices of silicon) in batches (\emph{lot}s).
Multiple circuits (\emph{dies}) are produced on a single wafer.
After the wafer's height profile is \emph{measured}, a light source \emph{exposes} the chip pattern onto a wafer through a projection mask (a \emph{reticle}).
A reticle may contain a full-sized pattern (\emph{full field}) or a smaller one (\emph{narrow field}).
Computational lithography software uses the \emph{measurements} to compensate for nano-scale imperfections during \emph{exposure}.

In this section the start of function call $f$ is denoted as $f^\uparrow$ and its end as $f^\downarrow$.

\vspace{-6pt}
\subsection{Case Study 1: Legacy Component Technology Migration}
\vspace{-2pt}

For the first case study, we look at a relatively small computational lithography component, developed and maintained by two engineers.
It is internally implemented using legacy end-of-life technology and is migrated to new technology, without changes to its external interface.
The engineers thus expect to see the same external behavior in communications with the other components, and we apply our approach to see whether this is indeed the case.

We observe six executions, using three different test sets for both the legacy and new implementations.
The \emph{integration} test set contains integration tests.
The \emph{overruling} and \emph{verification} test sets both test different configuration options and functionality of the component.
Each test set contains multiple tests.
For reasons of confidentially we do not explain the test sets in more detail.

For each observed execution, we obtain an execution log capturing the component's runtime communications with other components.
The log for each execution is split into separate logs for each of the functions in the component's external interface.
We use model learning~\cite{Hooimeijer2022} to obtain six model sets (one for each execution), with 11 interface functions of the component as entities.
The model sets together contain 46 models with behavior, with 2 to 578 states per model, and a sum total of 1,330 states.

\begin{figure}[!t]
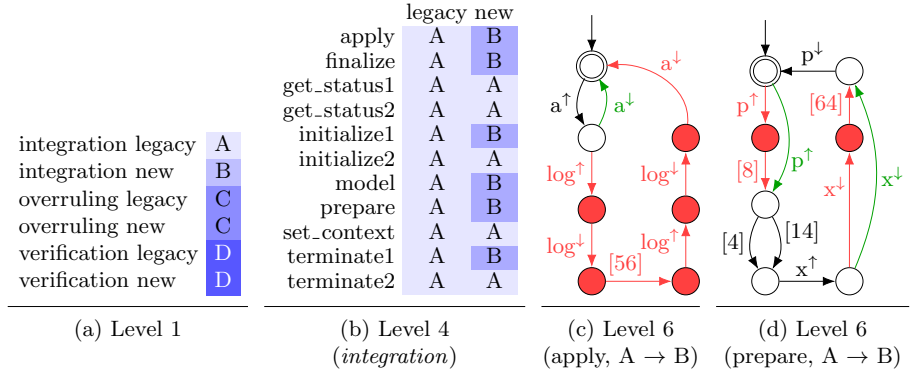

    \vspace{-20pt}
    \centering
    \captionsetup[subfigure]{justification=centering}
    \resizebox{\textwidth}{!}{%
        \subfloat[Level 1]{
            \begin{tikzpicture}[scale=1.5]
                \tikzset{
  ->,>=stealth',auto,node distance=22pt,
  arrows={-{Latex[scale=1.05]}},
  every initial by arrow/.style={-{Latex[scale=1.05]}},
  every initial by arrow/.append style={anchor/.append style={shape=coordinate}},
  every node/.style={font=\small},
  every state/.style={circle, draw, minimum size=11pt, initial text=},
  initial distance=10pt,
  initial where=left,
  accepting/.style={double, double distance=1pt},
  background rectangle/.style={draw=red, top color=blue, rounded corners},
  obsvariant/.style ={state,ellipse,         fill=black!10,inner sep=0pt,minimum size=0pt,align=center,font=\footnotesize},
  calcvariant/.style={state,diamond,aspect=2,fill=black!60,inner sep=0pt,minimum size=0pt,align=center,font=\footnotesize,text=white},
  myedge/.style={}
}

\node[] {
    \begingroup
    \begin{tabular}{ l *{1}{>{\centering\arraybackslash}m{1.2em}} }
        integration legacy\,  & \cellcolor{blue!10}A              \\
        integration new\,     & \cellcolor{blue!25}B              \\
        overruling legacy\,   & \cellcolor{blue!45}C              \\
        overruling new\,      & \cellcolor{blue!45}C              \\
        verification legacy\, & \cellcolor{blue!66}\color{white}D \\
        verification new\,    & \cellcolor{blue!66}\color{white}D \\
    \end{tabular}
    \endgroup
};

\draw[-] [black] ($ (current bounding box.south west) + (0pt,-1pt) $) to (current bounding box.south east);

            \end{tikzpicture}
            \label{fig:4-case-study-1a-lvl1}
        }
        \subfloat[Level 4\\(\emph{integration})]{
            \begin{tikzpicture}[scale=1.5]
                \tikzset{
  ->,>=stealth',auto,node distance=22pt,
  arrows={-{Latex[scale=1.05]}},
  every initial by arrow/.style={-{Latex[scale=1.05]}},
  every initial by arrow/.append style={anchor/.append style={shape=coordinate}},
  every node/.style={font=\small},
  every state/.style={circle, draw, minimum size=11pt, initial text=},
  initial distance=10pt,
  initial where=left,
  accepting/.style={double, double distance=1pt},
  background rectangle/.style={draw=red, top color=blue, rounded corners},
  obsvariant/.style ={state,ellipse,         fill=black!10,inner sep=0pt,minimum size=0pt,align=center,font=\footnotesize},
  calcvariant/.style={state,diamond,aspect=2,fill=black!60,inner sep=0pt,minimum size=0pt,align=center,font=\footnotesize,text=white},
  myedge/.style={}
}

\node[] {
    \begingroup
    \renewcommand{\arraystretch}{0.9}
    \begin{tabular}{ r c c }
        ~                           & legacy                   & new                   \\
        apply\,                     & \cellcolor{blue!10}A     & \cellcolor{blue!33}B  \\
        finalize\,                  & \cellcolor{blue!10}A     & \cellcolor{blue!33}B  \\
        get\_\hspace{.1em}status1\, & \cellcolor{blue!10}A     & \cellcolor{blue!10}A  \\
        get\_\hspace{.1em}status2\, & \cellcolor{blue!10}A     & \cellcolor{blue!10}A  \\
        initialize1\,               & \cellcolor{blue!10}A     & \cellcolor{blue!33}B  \\
        initialize2\,               & \cellcolor{blue!10}A     & \cellcolor{blue!10}A  \\
        model\,                     & \cellcolor{blue!10}A     & \cellcolor{blue!33}B  \\
        prepare\,                   & \cellcolor{blue!10}A     & \cellcolor{blue!33}B  \\
        set\_\hspace{.1em}context\, & \cellcolor{blue!10}A     & \cellcolor{blue!10}A  \\
        terminate1\,                & \cellcolor{blue!10}A     & \cellcolor{blue!33}B  \\
        terminate2\,                & \cellcolor{blue!10}A     & \cellcolor{blue!10}A  \\
    \end{tabular}
    \endgroup
};

\draw[-] [black] ($ (current bounding box.south west) + (0pt,-1pt) $) to (current bounding box.south east);

            \end{tikzpicture}
            \label{fig:4-case-study-1a-lvl4}
        }
        \subfloat[Level 6\\(apply, A $\rightarrow$ B)]{
            \begin{tikzpicture}[scale=1.5]
                \input{figures/4-case-study-1a-lvl6-apply.tex}
            \end{tikzpicture}
            \label{fig:4-case-study-1a-lvl6-apply}
        }
        \subfloat[Level 6\\(prepare, A $\rightarrow$ B)]{
            \begin{tikzpicture}[scale=1.5]
                \input{figures/4-case-study-1a-lvl6-prepare.tex}
            \end{tikzpicture}
            \label{fig:4-case-study-1a-lvl6-prepare}
        }
    }
    \caption{First results for case study 1: complete level 1, level 4 for the \emph{integration} test set, and level 6 with variants A vs B for functions `apply' and `prepare'.}
    \label{fig:4-case-study-1a}
    \vspace{-10pt}
\end{figure}

We discuss the results of applying our approach, per level.

\smallskip

\textbf{Level 1} (Figure~\ref{fig:4-case-study-1a-lvl1}):
Only for \emph{integration} there are differences in behavior between the legacy and new implementations.
As the other two test sets show no differences, they do not need further inspection.
Given that we then have only two model sets left, we skip levels 2 and 3, and proceed directly to level 4.

\textbf{Level 4} (Figure~\ref{fig:4-case-study-1a-lvl4}):
We see the 11 functions, anonymized for confidentiality reasons, and their behavioral variants.
Only 6 out of 11 entities show differences in behavior, to be inspected in more detail.
Given that they all have only two variants per entity, we skip level 5 and proceed directly to level 6.

\textbf{Level 6} (Figures~\ref{fig:4-case-study-1a-lvl6-apply} and~\ref{fig:4-case-study-1a-lvl6-prepare}):
Figure~\ref{fig:4-case-study-1a-lvl6-apply} shows the diff NFA for function `apply' (abbreviated to `a'), for variant A to variant B.
The figure shows that the new implementation involves only the start and end of this function.
The legacy implementation has more behavior, as within the `apply' function it has 30 calls (with returns) to a `log' function.
In the figure, only the first and last of these calls (with their returns) are shown, and the remaining sequence of 56 transitions, representing 28 calls and their returns, is abbreviated to `$[56]$'.
Figure~\ref{fig:4-case-study-1a-lvl6-prepare} shows the diff NFA for function `prepare' (abbreviated to `p'), for variant A to variant B.
For reasons of confidentiality and presentational clarity again several sequences of transitions are abbreviated.
Here, the figure shows that the legacy implementation invokes the `log' function 4 and 32 times, indicated as `$[8]$' and `$[64]$', respectively, while the new implementation does not.

\smallskip

Having inspected the differences for only two entities, it appears that all `log' function calls are missing in the new implementation.
The component engineers confirmed that indeed for the new implementation the component was not yet hooked up to the logging framework.
Our approach clearly shows this regression.

To look for other differences in behavior, we remove all `log' function calls and returns from the models of the legacy implementation.
To do so, we rename all `log' function call and return events to $\varepsilon$ and apply weak-language normalization~\cite{Sipser2013}.
We then apply our approach again.

\smallskip

\textbf{Level 1} (Figure~\ref{fig:4-case-study-1b}): Looking at the new results for level 1, we immediately see that there are no more observed differences in behavior for the legacy and new implementations, for all three test sets.
We don't see any further regressions in behavior, and we don't have to go to further levels.

\smallskip

\begin{wrapfigure}{r}{0.31\textwidth}
    \vspace{-24pt}
    \begin{center}
        \begin{tikzpicture}[scale=4]
            \hspace{-2pt}
            \tikzset{
  ->,>=stealth',auto,node distance=22pt,
  arrows={-{Latex[scale=1.05]}},
  every initial by arrow/.style={-{Latex[scale=1.05]}},
  every initial by arrow/.append style={anchor/.append style={shape=coordinate}},
  every node/.style={font=\small},
  every state/.style={circle, draw, minimum size=11pt, initial text=},
  initial distance=10pt,
  initial where=left,
  accepting/.style={double, double distance=1pt},
  background rectangle/.style={draw=red, top color=blue, rounded corners},
  obsvariant/.style ={state,ellipse,         fill=black!10,inner sep=0pt,minimum size=0pt,align=center,font=\footnotesize},
  calcvariant/.style={state,diamond,aspect=2,fill=black!60,inner sep=0pt,minimum size=0pt,align=center,font=\footnotesize,text=white},
  myedge/.style={}
}

\node[] {
    \begingroup
    \begin{tabular}{ l *{1}{>{\centering\arraybackslash}m{1.2em}} }
        integration legacy\,  & \cellcolor{blue!10}A              \\
        integration new\,     & \cellcolor{blue!10}A              \\
        overruling legacy\,   & \cellcolor{blue!33}B              \\
        overruling new\,      & \cellcolor{blue!33}B              \\
        verification legacy\, & \cellcolor{blue!66}\color{white}C \\
        verification new\,    & \cellcolor{blue!66}\color{white}C \\
    \end{tabular}
    \endgroup
};

        \end{tikzpicture}
    \end{center}
    \vspace{-20pt}
    \caption{New results for case study 1: level 1.}
    \label{fig:4-case-study-1b}
    \vspace{-20pt}
\end{wrapfigure}

Given that this component has quite a good test set with adequate coverage, our approach is applied as an extra safety net that complements traditional testing, akin to differential testing~\cite{Gulzar2019}.
As any change in the (order of) communications with other components will show up in our models and comparisons, it is like having assertions for all external communications.
Both engineers find this valuable.
They would like to apply our methodology also for larger and more complex technology migrations, where they foresee even more value.

\subsection{Case Study 2: Test Coverage}

The second case study considers again the same component and three test sets from the first case study, but from a difference angle.
Instead of comparing the legacy and new implementation, we compare the three test sets against each other.
The goal is to see how the behaviors of the different test sets differ, and whether one or more test sets are perhaps superfluous.
We use the versions of the input models from the first case study where the `log' function is completely removed.
We discuss the results of applying our methodology, per level:

\begin{figure}[!b]
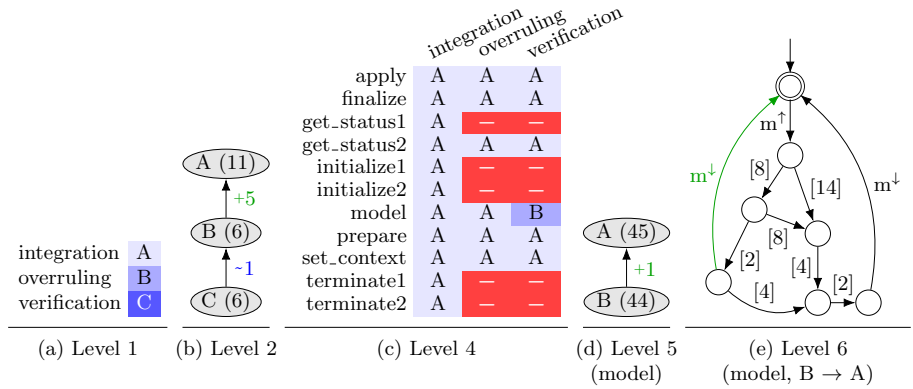

    \vspace{-20pt}
    \centering
    \captionsetup[subfigure]{justification=centering}
    \resizebox{\textwidth}{!}{%
        \subfloat[Level 1]{
            \begin{tikzpicture}[scale=1.5]
                \tikzset{
  ->,>=stealth',auto,node distance=22pt,
  arrows={-{Latex[scale=1.05]}},
  every initial by arrow/.style={-{Latex[scale=1.05]}},
  every initial by arrow/.append style={anchor/.append style={shape=coordinate}},
  every node/.style={font=\small},
  every state/.style={circle, draw, minimum size=11pt, initial text=},
  initial distance=10pt,
  initial where=left,
  accepting/.style={double, double distance=1pt},
  background rectangle/.style={draw=red, top color=blue, rounded corners},
  obsvariant/.style ={state,ellipse,         fill=black!10,inner sep=0pt,minimum size=0pt,align=center,font=\footnotesize},
  calcvariant/.style={state,diamond,aspect=2,fill=black!60,inner sep=0pt,minimum size=0pt,align=center,font=\footnotesize,text=white},
  myedge/.style={}
}

\node[] {
    \begingroup
    \begin{tabular}{ l *{1}{>{\centering\arraybackslash}m{1.2em}} }
        integration\,  & \cellcolor{blue!10}A              \\
        overruling\,   & \cellcolor{blue!33}B              \\
        verification\, & \cellcolor{blue!66}\color{white}C \\
    \end{tabular}
    \endgroup
};

\draw[-] [black] ($ (current bounding box.south west) + (0pt,-1pt) $) to (current bounding box.south east);

            \end{tikzpicture}
            \label{fig:4-case-study-2-lvl1}
        }
        \subfloat[Level 2]{
            \begin{tikzpicture}[scale=1.5]
                \input{figures/4-case-study-2-lvl2.tex}
            \end{tikzpicture}
            \label{fig:4-case-study-2-lvl2}
        }
        \subfloat[Level 4]{
            \begin{tikzpicture}[scale=1.5]
                \tikzset{
  ->,>=stealth',auto,node distance=22pt,
  arrows={-{Latex[scale=1.05]}},
  every initial by arrow/.style={-{Latex[scale=1.05]}},
  every initial by arrow/.append style={anchor/.append style={shape=coordinate}},
  every node/.style={font=\small},
  every state/.style={circle, draw, minimum size=11pt, initial text=},
  initial distance=10pt,
  initial where=left,
  accepting/.style={double, double distance=1pt},
  background rectangle/.style={draw=red, top color=blue, rounded corners},
  obsvariant/.style ={state,ellipse,         fill=black!10,inner sep=0pt,minimum size=0pt,align=center,font=\footnotesize},
  calcvariant/.style={state,diamond,aspect=2,fill=black!60,inner sep=0pt,minimum size=0pt,align=center,font=\footnotesize,text=white},
  myedge/.style={}
}

\node[] {
    \begingroup
    \renewcommand{\arraystretch}{0.9}
    \begin{tabular}{ r *{3}{>{\centering\arraybackslash}m{2em}}  }
        ~                           & ~~\rotatebox{25}{integration} & ~~\rotatebox{25}{overruling}                   & ~~\rotatebox{25}{verification}                 \\
        apply\,                     & \cellcolor{blue!10}A          & \cellcolor{blue!10}A                           & \cellcolor{blue!10}A                           \\
        finalize\,                  & \cellcolor{blue!10}A          & \cellcolor{blue!10}A                           & \cellcolor{blue!10}A                           \\
        get\_\hspace{.1em}status1\, & \cellcolor{blue!10}A          & \cellcolor{diff-red}\color{white}$\diffabsent$ & \cellcolor{diff-red}\color{white}$\diffabsent$ \\
        get\_\hspace{.1em}status2\, & \cellcolor{blue!10}A          & \cellcolor{blue!10}A                           & \cellcolor{blue!10}A                           \\
        initialize1\,               & \cellcolor{blue!10}A          & \cellcolor{diff-red}\color{white}$\diffabsent$ & \cellcolor{diff-red}\color{white}$\diffabsent$ \\
        initialize2\,               & \cellcolor{blue!10}A          & \cellcolor{diff-red}\color{white}$\diffabsent$ & \cellcolor{diff-red}\color{white}$\diffabsent$ \\
        model\,                     & \cellcolor{blue!10}A          & \cellcolor{blue!10}A                           & \cellcolor{blue!33}B                           \\
        prepare\,                   & \cellcolor{blue!10}A          & \cellcolor{blue!10}A                           & \cellcolor{blue!10}A                           \\
        set\_\hspace{.1em}context\, & \cellcolor{blue!10}A          & \cellcolor{blue!10}A                           & \cellcolor{blue!10}A                           \\
        terminate1\,                & \cellcolor{blue!10}A          & \cellcolor{diff-red}\color{white}$\diffabsent$ & \cellcolor{diff-red}\color{white}$\diffabsent$ \\
        terminate2\,                & \cellcolor{blue!10}A          & \cellcolor{diff-red}\color{white}$\diffabsent$ & \cellcolor{diff-red}\color{white}$\diffabsent$ \\
    \end{tabular}
    \endgroup
};

\draw[-] [black] ($ (current bounding box.south west) + (0pt,-1pt) $) to (current bounding box.south east);

            \end{tikzpicture}
            \label{fig:4-case-study-2-lvl4}
        }
        \subfloat[Level 5\\(model)]{
            \begin{tikzpicture}[scale=1.5]
                \input{figures/4-case-study-2-lvl5.tex}
            \end{tikzpicture}
            \label{fig:4-case-study-2-lvl5}
        }
        \subfloat[Level 6\\(model, B $\rightarrow$ A)]{
            \begin{tikzpicture}[scale=1.5]
                \input{figures/4-case-study-2-lvl6.tex}
            \end{tikzpicture}
            \label{fig:4-case-study-2-lvl6}
        }
    }
    \vspace{-11pt}
    \caption{Results for case study 2: complete levels 1, 2 and 4, level 5 for function `model', and level 6 for function `model' variants B vs A.}
    \label{fig:4-case-study-2}
\end{figure}

\smallskip

\textbf{Level 1} (Figure~\ref{fig:4-case-study-2-lvl1}):
The three tests sets have different behavior (A\,--\,C).

\textbf{Level 2} (Figure~\ref{fig:4-case-study-2-lvl2}):
The \emph{integration} test set (variant A) has behavior for all 11 functions, and the other two test sets (B, C) for 5 fewer functions, i.e., 6 functions.
Also, \emph{integration} (A) includes all the behavior of the other two test sets, while \emph{verification} (C) differs from \emph{overruling} (B) by only one function.
As all variants are (transitively) related in the lattice, we skip level 3.

\textbf{Level 4} (Figure~\ref{fig:4-case-study-2-lvl4}):
We clearly see which 5 functions are only used during the \emph{integration} tests.
The component engineers expect this difference, as for \emph{overruling} and \emph{verification} these 5 functions are stubbed internally and are thus not externally visible.
Also, for the \emph{verification} tests only the `model' function has different behavior.
We inspect this further in level 5.

\textbf{Level 5} (Figure~\ref{fig:4-case-study-2-lvl5}):
The behavior of the `model' function for variant B (\emph{verification}) is included in that of variant A (\emph{integration} and \emph{overruling}), which has one additional transition.
We inspect this further in level 6.

\textbf{Level 6} (Figure~\ref{fig:4-case-study-2-lvl6}):
Here we see the diff NFA for function `model' (abbreviated to `m'), for variant B to variant A, following the arrow in the level 5 lattice.
For confidentiality and presentational clarity we annotate other arrows with $[n]$ to abbreviate $n$ transitions in sequence.
The one extra transition of function variant A is clearly visible.
There it is possible to return to the initial state earlier on, skipping part of the behavior of the state machine.
The engineers again expect this, as some functionality is not activated depending on the component configuration.

\smallskip

The comparison results suggest that since the \emph{integration} test set covers more behavior than the other two test sets, those other two test sets can be removed.
This would be a valid conclusion, if one only considers function call order, as we do for our methodology.
However, functions could have different behavior for different arguments.
If this results in a difference in which functions are called or in what order they are called, then our approach will highlight such differences.
If however for different configurations there are differences in which paths though a state machine are taken for which argument values, while each path is still taken for some argument value, this would not be visible with our current approach.
The different test sets that we consider do indeed test different configurations using different argument values, and hence they do add value and can not simply be removed.
Fully taking the influence of argument values into account is considered future work.

In any regard, our methodology provides insight into the behavioral differences for the various configurations and functional scenarios considered by the different test sets.
This can be automatically obtained even by engineers who are not domain experts.

\subsection{Case Study 3: System Behavior Matching Recipe}

For the third case study, we investigate how \emph{recipes} containing information on the number of wafers and used reticles relate to the system behavior.
ASML's customers can specify their own recipes to configure their lithography systems for their purposes, e.g., to create CPU or memory chips.
The software running on the systems will exhibit different behavior for different recipes, and thus software behavior is a lens to look at system behavior.

\begin{figure}[!b]
    \centering
    \begingroup
    \setlength{\tabcolsep}{8pt}
    \def\arraystretch{1.2}
    \small
    \begin{tabular}{l c c c c c c}
        \textbf{~}       & \textbf{Lot 1} & \textbf{Lot 2} & \textbf{Lot 3} & \textbf{Lot 4} & \textbf{Lot 5} & \textbf{Lot 6} \\
        \textbf{Wafers}  & 5              & 5              & 15             & 15             & 15             & 15             \\
        \textbf{Reticle} & 96*X           & 96*Y           & 96*X           & 96*Y           & 124*X, 1*Y     & 125*X          \\
        \textbf{Field}   & Full           & Full           & Full           & Full           & Narrow         & Narrow         \\
    \end{tabular}
    \endgroup
    \caption{Case study 3: recipes for the different lots.}
    \label{fig:4-recipe}
\end{figure}

Figure~\ref{fig:4-recipe} shows the recipes that we consider for this case study.
For reasons of confidentially, we don't explain the origin of these recipes and we consider only the details relevant for this case study.
There are six lots, each with their own recipe.
Lots 1 and 2 have five wafers each and the other lots have 15 wafers each.
There are two reticles, X and Y.
For lot 1, reticle X is used 96 times, one for each die.
Lot 5 uses both reticles.
Exposure can be done using full field or narrow field, where narrow field leads to more exposures (125 rather than 96).

We consider the behavior of the exposure sub-system, i.e., 32 software components involved in the high-level exposure control.
Observing the system execution for about an hour as it initializes and processes lots, we obtain a single execution log capturing all observed inter-component communications.
This log is split into multiple logs, one for each of the 85 exposures (one per wafer and for lot 5 twice per wafer as it uses two reticles).
The exposure logs are further split into separate logs for each of the components, containing only their interactions with the other components.
We use model learning~\cite{Hooimeijer2022} to obtain 85 model sets (one per exposure), containing models of the 32 components (entities).
Model sets may lack a certain component model if that component did not interact with other components during the corresponding exposure.
Figure~\ref{fig:4-case-study-3-sizes} shows the sizes of the input models in number of states.
The 85 model sets together contain 2,386 models with behavior, with 2 to 7,070 states per model, and a sum total of 495,505 states, making this a large case study.

\begin{figure}[!t]
    \centering
    \begin{tikzpicture}[scale=1]
        \input{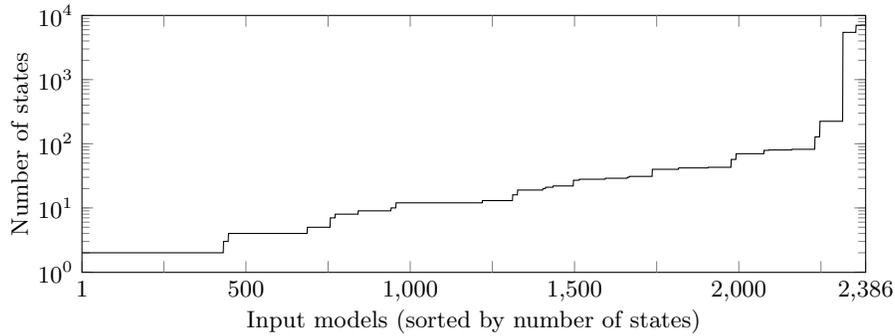}
    \end{tikzpicture}
    \vspace{-10pt}
    \caption{Case study 3: sizes of the input models with behavior.}
    \label{fig:4-case-study-3-sizes}
\end{figure}

We apply our methodology and discuss the results level by level, skipping levels 2 and 5 as they are less relevant for this case study.

\begin{figure}[!b]
    \centering
    \begin{tikzpicture}[scale=1.5]
        \tikzset{
  ->,>=stealth',auto,node distance=22pt,
  arrows={-{Latex[scale=1.05]}},
  every initial by arrow/.style={-{Latex[scale=1.05]}},
  every initial by arrow/.append style={anchor/.append style={shape=coordinate}},
  every node/.style={font=\small},
  every state/.style={circle, draw, minimum size=11pt, initial text=},
  initial distance=10pt,
  initial where=left,
  accepting/.style={double, double distance=1pt},
  background rectangle/.style={draw=red, top color=blue, rounded corners},
  obsvariant/.style ={state,ellipse,         fill=black!10,inner sep=0pt,minimum size=0pt,align=center,font=\footnotesize},
  calcvariant/.style={state,diamond,aspect=2,fill=black!60,inner sep=0pt,minimum size=0pt,align=center,font=\footnotesize,text=white},
  myedge/.style={}
}

\node[] (t1) {
    \begingroup
    \setlength{\tabcolsep}{3pt}
    \begin{tabular}{ l c }
        1-1   & \cellcolor{blue!11}A \\
        1-2   & \cellcolor{blue!25}B \\
        1-3   & \cellcolor{blue!25}B \\
        1-4   & \cellcolor{blue!25}B \\
        1-5   & \cellcolor{blue!25}B \\

        ~     & ~ \\

        2-1   & \cellcolor{blue!11}A \\
        2-2   & \cellcolor{blue!25}B \\
        2-3   & \cellcolor{blue!25}B \\
        2-4   & \cellcolor{blue!55}\color{white}C \\
        2-5   & \cellcolor{blue!25}B \\
    \end{tabular}
    \endgroup
};

\node[right of=t1,xshift=35pt] (t2) {
    \begingroup
    \setlength{\tabcolsep}{3pt}
    \begin{tabular}{ l c }
        3-1   & \cellcolor{blue!11}A \\
        3-2   & \cellcolor{blue!25}B \\
        3-3   & \cellcolor{blue!25}B \\
        3-4   & \cellcolor{blue!25}B \\
        3-5   & \cellcolor{blue!25}B \\
        3-6   & \cellcolor{blue!25}B \\
        3-7   & \cellcolor{blue!25}B \\
        3-8   & \cellcolor{blue!25}B \\
        3-9   & \cellcolor{blue!25}B \\
        3-10  & \cellcolor{blue!25}B \\
        3-11  & \cellcolor{blue!55}\color{white}C \\
        3-12  & \cellcolor{blue!25}B \\
        3-13  & \cellcolor{blue!25}B \\
        3-14  & \cellcolor{blue!25}B \\
        3-15  & \cellcolor{blue!25}B \\
    \end{tabular}
    \endgroup
};

\node[right of=t2,xshift=35pt] (t3) {
    \begingroup
    \setlength{\tabcolsep}{3pt}
    \begin{tabular}{ l c }
        4-1   & \cellcolor{blue!70}\color{white}D \\
        4-2   & \cellcolor{blue!25}B \\
        4-3   & \cellcolor{blue!25}B \\
        4-4   & \cellcolor{blue!25}B \\
        4-5   & \cellcolor{blue!25}B \\
        4-6   & \cellcolor{blue!25}B \\
        4-7   & \cellcolor{blue!25}B \\
        4-8   & \cellcolor{blue!25}B \\
        4-9   & \cellcolor{blue!25}B \\
        4-10  & \cellcolor{blue!25}B \\
        4-11  & \cellcolor{blue!25}B \\
        4-12  & \cellcolor{blue!25}B \\
        4-13  & \cellcolor{blue!90}\color{white}E \\
        4-14  & \cellcolor{blue!25}B \\
        4-15  & \cellcolor{blue!25}B \\
    \end{tabular}
    \endgroup
};

\node[right of=t3,xshift=35pt] (t4) {
    \begingroup
    \setlength{\tabcolsep}{3pt}
    \begin{tabular}{ l c }
        5-1A  & \cellcolor{black!14!yellow}F \\
        5-1B  & \cellcolor{black!7!yellow}G \\
        5-2B  & \cellcolor{yellow}H \\
        5-2A  & \cellcolor{red!15!yellow}I \\
        5-3A  & \cellcolor{red!30!yellow}J \\
        5-3B  & \cellcolor{black!7!yellow}G \\
        5-4B  & \cellcolor{yellow}H \\
        5-4A  & \cellcolor{red!15!yellow}I \\
        5-5A  & \cellcolor{red!30!yellow}J \\
        5-5B  & \cellcolor{black!7!yellow}G \\
        5-6B  & \cellcolor{yellow}H \\
        5-6A  & \cellcolor{red!15!yellow}I \\
        5-7A  & \cellcolor{red!60!yellow}\color{white}K \\
        5-7B  & \cellcolor{black!7!yellow}G \\
        5-8B  & \cellcolor{yellow}H \\
        5-8A  & \cellcolor{red!15!yellow}I \\
    \end{tabular}
    \endgroup
};

\node[right of=t4,xshift=35pt] (t5) {
    \begingroup
    \setlength{\tabcolsep}{3pt}
    \begin{tabular}{ l c }
        5-9A  & \cellcolor{red!60!yellow}\color{white}K \\
        5-9B  & \cellcolor{black!7!yellow}G \\
        5-10B & \cellcolor{yellow}H \\
        5-10A & \cellcolor{red!15!yellow}I \\
        5-11A & \cellcolor{red!60!yellow}\color{white}K \\
        5-11B & \cellcolor{black!7!yellow}G \\
        5-12B & \cellcolor{yellow}H \\
        5-12A & \cellcolor{red!15!yellow}I \\
        5-13A & \cellcolor{red!60!yellow}\color{white}K \\
        5-13B & \cellcolor{red!90!yellow}\color{white}L \\
        5-14B & \cellcolor{yellow}H \\
        5-14A & \cellcolor{red!15!yellow}I \\
        5-15A & \cellcolor{red!60!yellow}\color{white}K \\
        5-15B & \cellcolor{red!90!yellow}\color{white}L \\
    \end{tabular}
    \endgroup
};

\node[right of=t5,xshift=35pt] (t6) {
    \begingroup
    \setlength{\tabcolsep}{3pt}
    \begin{tabular}{ l c }
        6-1  & \cellcolor{green!55}M \\
        6-2  & \cellcolor{black!11!green}N \\
        6-3  & \cellcolor{black!11!green}N \\
        6-4  & \cellcolor{black!11!green}N \\
        6-5  & \cellcolor{black!11!green}N \\
        6-6  & \cellcolor{black!11!green}N \\
        6-7  & \cellcolor{black!11!green}N \\
        6-8  & \cellcolor{black!11!green}N \\
        6-9  & \cellcolor{black!33!green}\color{white}O \\
        6-10 & \cellcolor{black!11!green}N \\
        6-11 & \cellcolor{black!11!green}N \\
        6-12 & \cellcolor{black!11!green}N \\
        6-13 & \cellcolor{black!11!green}N \\
        6-14 & \cellcolor{black!11!green}N \\
        6-15 & \cellcolor{black!33!green}\color{white}O \\
    \end{tabular}
    \endgroup
};

    \end{tikzpicture}
    \vspace{-8pt}
    \caption{Results for case study 3: level 1.}
    \label{fig:4-case-study-3-lvl1}
    \vspace{-12pt}
\end{figure}

\vspace{-4pt}
\vspace{\baselineskip}
\textbf{Level 1} (Figure~\ref{fig:4-case-study-3-lvl1}):
We discuss multiple observations based on patterns that are visible in level~1.
Different gradient colors are used for presentational clarity.

\begin{enumerate}[a)]
    \item \emph{First exposure of a lot}:
    For lots 1\,--\,4, the main behavior variant is variant B.
    The first exposures of these lots however all have different behavior (A, D).

    \item \emph{Changes during a lot}:
    For lots 2\,--\,4 we also see different behavior for some exposures later during the lot (C, E).

    \item \emph{Reticle swaps}:
    All exposures of lots 5 (F\,--\,L) have behavior different than the other lots (A\,--\,E, M\,--\,O).
    Lot 5 is the only lot where two reticles are used per wafer, and thus reticles must be swapped regularly.
    To minimize the number of swaps, the system uses an `XYYX' pattern for every two wafers (first wafer reticle `X', first wafer reticle `Y', second wafer reticle `Y', second wafer reticle `X').
    These patterns of four exposures are clearly visible in the model set variants (J\,--\,G\,--\,H\,--\,I, K\,--\,G\,--\,H\,--\,I).

    \item \emph{Full field vs narrow field}:
    The difference between lots 1 and 3 compared to lot 6 is the use of full vs narrow field.
    The behavior for lots 1 and 3 (A\,--\,C) and lot 6 (M\,--\,O) differ, but they have similar structure (mostly the same variant, first exposure and some exposures during the lot are different).
\end{enumerate}

\begin{figure}[!p]
    \centering
    \makebox[\textwidth][c]{
        \resizebox{1.6\textwidth}{!}{%
            \input{figures/4-case-study-3-lvl3.tex}
        }
    }
    \caption{Results for case study 3: level 3.}
    \label{fig:4-case-study-3-lvl3}
\end{figure}

\begin{figure}[!p]
    \centering
    \resizebox{\textwidth}{!}{%
        \rotatebox{90}{%
            \begingroup
\renewcommand{\arraystretch}{1.5}
\aboverulesep = 0mm
\belowrulesep = 0mm
\arrayrulecolor{black}
\begin{tabular}{ l >{\centering}m{.8em} >{\centering}m{.8em} >{\centering}m{.8em} >{\centering}m{.8em} >{\centering}m{.8em} | >{\centering}m{.8em} >{\centering}m{.8em} >{\centering}m{.8em} >{\centering}m{.8em} >{\centering}m{.8em} | >{\centering}m{.8em} >{\centering}m{.8em} >{\centering}m{.8em} >{\centering}m{.8em} >{\centering}m{.8em} >{\centering}m{.8em} >{\centering}m{.8em} >{\centering}m{.8em} >{\centering}m{.8em} >{\centering}m{.8em} >{\centering}m{.8em} >{\centering}m{.8em} >{\centering}m{.8em} >{\centering}m{.8em} >{\centering}m{.8em} | >{\centering}m{.8em} >{\centering}m{.8em} >{\centering}m{.8em} >{\centering}m{.8em} >{\centering}m{.8em} >{\centering}m{.8em} >{\centering}m{.8em} >{\centering}m{.8em} >{\centering}m{.8em} >{\centering}m{.8em} >{\centering}m{.8em} >{\centering}m{.8em} >{\centering}m{.8em} >{\centering}m{.8em} >{\centering}m{.8em} | >{\centering}m{.8em} >{\centering}m{.8em} >{\centering}m{.8em} >{\centering}m{.8em} >{\centering}m{.8em} >{\centering}m{.8em} >{\centering}m{.8em} >{\centering}m{.8em} >{\centering}m{.8em} >{\centering}m{.8em} >{\centering}m{.8em} >{\centering}m{.8em} >{\centering}m{.8em} >{\centering}m{.8em} >{\centering}m{.8em} >{\centering}m{.8em} >{\centering}m{.8em} >{\centering}m{.8em} >{\centering}m{.8em} >{\centering}m{.8em} >{\centering}m{.8em} >{\centering}m{.8em} >{\centering}m{.8em} >{\centering}m{.8em} >{\centering}m{.8em} >{\centering}m{.8em} >{\centering}m{.8em} >{\centering}m{.8em} >{\centering}m{.8em} >{\centering}m{.8em} | >{\centering}m{.8em} >{\centering}m{.8em} >{\centering}m{.8em} >{\centering}m{.8em} >{\centering}m{.8em} >{\centering}m{.8em} >{\centering}m{.8em} >{\centering}m{.8em} >{\centering}m{.8em} >{\centering}m{.8em} >{\centering}m{.8em} >{\centering}m{.8em} >{\centering}m{.8em} >{\centering}m{.8em} >{\centering\arraybackslash}m{.8em} }
       & \rotatebox{90}{1-1}                            & \rotatebox{90}{1-2}                            & \rotatebox{90}{1-3}                            & \rotatebox{90}{1-4}                            & \rotatebox{90}{1-5}                            & \rotatebox{90}{2-1}                            & \rotatebox{90}{2-2}                            & \rotatebox{90}{2-3}                            & \rotatebox{90}{2-4}                            & \rotatebox{90}{2-5}                            & \rotatebox{90}{3-1}                            & \rotatebox{90}{3-2}                            & \rotatebox{90}{3-3}                            & \rotatebox{90}{3-4}                            & \rotatebox{90}{3-5}                            & \rotatebox{90}{3-6}                            & \rotatebox{90}{3-7}                            & \rotatebox{90}{3-8}                            & \rotatebox{90}{3-9}                            & \rotatebox{90}{3-10}                           & \rotatebox{90}{3-11}                           & \rotatebox{90}{3-12}                           & \rotatebox{90}{3-13}                           & \rotatebox{90}{3-14}                           & \rotatebox{90}{3-15}                           & \rotatebox{90}{4-1}                            & \rotatebox{90}{4-2}                            & \rotatebox{90}{4-3}                            & \rotatebox{90}{4-4}                            & \rotatebox{90}{4-5}                            & \rotatebox{90}{4-6}                            & \rotatebox{90}{4-7}                            & \rotatebox{90}{4-8}                            & \rotatebox{90}{4-9}                            & \rotatebox{90}{4-10}                           & \rotatebox{90}{4-11}                           & \rotatebox{90}{4-12}                           & \rotatebox{90}{4-13}                           & \rotatebox{90}{4-14}                           & \rotatebox{90}{4-15}                           & \rotatebox{90}{5-1X}              & \rotatebox{90}{5-1Y}                           & \rotatebox{90}{5-2Y}                           & \rotatebox{90}{5-2X}                           & \rotatebox{90}{5-3X}                           & \rotatebox{90}{5-3Y}                           & \rotatebox{90}{5-4Y}                           & \rotatebox{90}{5-4X}                           & \rotatebox{90}{5-5X}                           & \rotatebox{90}{5-5Y}                           & \rotatebox{90}{5-6Y}                           & \rotatebox{90}{5-6X}                           & \rotatebox{90}{5-7X}                           & \rotatebox{90}{5-7Y}                           & \rotatebox{90}{5-8Y}                           & \rotatebox{90}{5-8X}                           & \rotatebox{90}{5-9X}                           & \rotatebox{90}{5-9Y}                           & \rotatebox{90}{5-10Y}                          & \rotatebox{90}{5-10X}                          & \rotatebox{90}{5-11X}                          & \rotatebox{90}{5-11Y}                          & \rotatebox{90}{5-12Y}                          & \rotatebox{90}{5-12X}                          & \rotatebox{90}{5-13X}                          & \rotatebox{90}{5-13Y}                          & \rotatebox{90}{5-14Y}                          & \rotatebox{90}{5-14X}                          & \rotatebox{90}{5-15X}                          & \rotatebox{90}{5-15Y}                          & \rotatebox{90}{6-1}                            & \rotatebox{90}{6-2}                            & \rotatebox{90}{6-3}                            & \rotatebox{90}{6-4}                            & \rotatebox{90}{6-5}                            & \rotatebox{90}{6-6}                            & \rotatebox{90}{6-7}                            & \rotatebox{90}{6-8}                            & \rotatebox{90}{6-9}                            & \rotatebox{90}{6-10}                           & \rotatebox{90}{6-11}                           & \rotatebox{90}{6-12}                           & \rotatebox{90}{6-13}                           & \rotatebox{90}{6-14}                           & \rotatebox{90}{6-15}                           \\
\end{tabular}
\endgroup

        }
    }
    \caption{Results for case study 3: level 4. The component names have been anonymized for confidentially reasons.}
    \label{fig:4-case-study-3-lvl4}
\end{figure}

\textbf{Level 3} (Figure~\ref{fig:4-case-study-3-lvl3}):
We elaborate on each of the four observations using the results for level 3.

\begin{enumerate}[a)]
    \item \emph{First exposure of a lot}:
    For lots 1\,--\,4, we mainly see regular behavior (dark green, $0$ components with different behavior).
    For the first exposures of these lots we do see differences (yellow lines, mainly $2$ or $3$ components).

    \item \emph{Changes during a lot}:
    For lots 2\,--\,4 we again see differences for some exposures later during the lot (light green and yellow lines, mainly $1$ or $2$ components).

    \item \emph{Reticle swaps}:
    The reticle swaps are again very much visible for lot 5 (vertical orange, red and light green lines in a repeating pattern of 4 columns).

    \item \emph{Full field vs narrow field}:
    Observe the differences between thick-border enclosed areas left and right of the figure.
    These full field (lots 1\,+\,3) vs narrow field (lot 6) differences seem to be caused by a single component.
\end{enumerate}

\vspace{-3pt}
\textbf{Level 4} (Figure~\ref{fig:4-case-study-3-lvl4}):
The observations are detailed even further using the results for level 4.
\vspace{-3pt}

\begin{enumerate}[a)]
    \item \emph{First exposure of a lot}:
    The differences in first exposures of lots 1\,--\,4 can be attributed primarily to components C1 and C21, and for lot 4 also to C28.

    \item \emph{Changes during a lot}:
    The changes for exposures during lots 2\,--\,4 can be attributed to components C4, C9 and C28.

    \item \emph{Reticle swaps}:
    The reticle swap differences concern many components.
    For several components (e.g., C2, C6, C9) we again see the `XYYX' reticle swap pattern.
    For some other components (e.g., C3, C4) we see a `VWVW' pattern instead, relating to first vs second exposure of a wafer.

    \item \emph{Full field vs narrow field}:
    Indeed only one component (C9) causes the full field (lots 1\,+\,3) vs narrow field (lot 6) differences (variants A/B vs G).
\end{enumerate}

\begin{figure}[!t]
    \centering
    \begin{tikzpicture}[scale=1]
        \input{figures/4-case-study-3-lvl6.tex}
    \end{tikzpicture}
    \vspace{-6pt}
    \caption{Results for case study 3: excerpt of level 6 (C21, A $\rightarrow$ B).}
    \label{fig:4-case-study-3-lvl6}
    \vspace{-12pt}
\end{figure}

\textbf{Level 6} (Figure~\ref{fig:4-case-study-3-lvl6}):
For reasons of confidentiality, we focus only on the \emph{first exposure of a lot} differences.
We inspect level 6 for variants A and B of component C21.
Figure~\ref{fig:4-case-study-3-lvl6} shows a part of the diff state machine, with `l' a logging function, `i' a function to get some information, and `q' a query function.
For confidentiality reasons we don't explain the functions in more detail.
The upper and lower paths indicate that both versions can skip the calls to `q'.
The only difference is that variant A (first wafer, in red) calls `i' before calling `q', while variant B (other wafers, in green) does not.
The company's domain experts are well aware of such `first wafer effects'.

\smallskip

The system behavior differs between wafers, and by going through the levels of our methodology we obtain progressive insights into these behavioral differences and how they relate to the recipes.
This allows engineers to understand how different configurations influence the system behavior, e.g., which components are affected by reticle swaps or full field vs narrow field, and in what way they behave differently.
While the input contains a large number of state machines, with an even larger number of states, our methodology allows engineers to step by step zoom in on parts of this behavior, thus making it suitable to analyze this large system.

Our approach has many potential applications.
For instance, understanding how certain configurations affect the system behavior is key when changing the system behavior.
Junior engineers can understand the system and its configurations without having to rely on domain experts.
Domain experts can check whether their mental views conform to reality, and adapt their mental views if they turn out to be outdated or incomplete.
Furthermore, if certain configurations have no effect at all on the system behavior, they could be removed from the system to avoid having to consider them when changing the system.

\section{Conclusions and Future Work}
\label{sec:conclusins-and-future-work}

We contribute a novel multi-level methodology for behavioral comparison of software-intensive systems.
It integrates multiple existing complementary methods to automatically compare the behavior of state machines.
Our methodology takes advantage of their complementary nature in a novel way, using six levels with progressive detail to handle the complexity of large industrial systems.

Our qualitative exploratory field study suggests that our approach allows one to inspect the behavioral differences of large systems, and that it has practical value for
getting insight into system behavior for various configurations and scenarios, and preventing regressions.
However, a more rigorous and quantitative evaluation of our methodology is still needed.

Our work is generically applicable as it works on state machines, which are widely used and understood in both computer science and industry.
We plan to research the generality of our approach by also applying it at other companies with software-intensive systems that have suitable state machine models~\cite{Bera2021}, and make the company-internal prototype tool publicly available.

Other future work includes extensions beyond comparing NFAs to consider also Extended Finite Automata and Timed Automata as input to our approach, and adding actionable insights beyond merely behavioral differences to further support change impact analysis.
Our methodology could also be applied to different use cases such as diagnosis of unstable tests and field issues.

\paragraph{Acknowledgments}
\addcontentsline{toc}{section}{Acknowledgements}
The authors would like to thank ASML for making this work possible and for supporting it.

\bibliographystyle{splncs04}
\bibliography{main}

\end{document}